\documentclass[aps,prl,twocolumn,groupedaddress]{revtex4}

\usepackage{graphicx}

\begin{document}

\preprint{LPSC 06-008}

\title{Transverse-Momentum Resummation for Slepton-Pair Production at the
 LHC}

\author{G.\ Bozzi}
\author{B.\ Fuks}
\author{M.\ Klasen}
\affiliation{Laboratoire de Physique Subatomique et de Cosmologie,
 Universit\'e Joseph Fourier/CNRS-IN2P3, 53 Avenue des Martyrs,
 F-38026 Grenoble, France}

\date{\today}

\begin{abstract}
 We perform a first precision calculation of the transverse-momentum ($q_T$)
 distribution of slepton pair and slepton-sneutrino associated production at
 the CERN Large Hadron Collider (LHC). We implement soft-gluon resummation
 at the next-to-leading logarithmic (NLL) level and consistently match the
 obtained result to the pure fixed-order perturbative result at leading
 order (LO) in the QCD coupling constant, {\em i.e.} ${\cal O}(\alpha_s)$.
 We give numerical predictions for $\tilde{\tau}_1 \tilde{\tau}_1^*$ and
 $\tilde{\tau}_1\tilde{\nu}_\tau^*+\tilde{\tau}_1^*\tilde{\nu}_\tau$
 production, also implementing recent parameterizations of non-perturbative
 effects. The results show a relevant contribution of resummation both in
 the small and intermediate $q_T$-regions and little dependence on
 unphysical scales and non-perturbative contributions.
\end{abstract}


\maketitle

\section{\label{introduction} Introduction}

The Minimal Supersymmetric Standard Model (MSSM) \cite{Nilles:1983ge,%
Haber:1984rc} is one of the most promising extensions of the Standard Model
(SM) of particle physics. It postulates a symmetry between fermionic and
bosonic degrees of freedom in nature and predicts the existence of a
fermionic (bosonic) supersymmetric (SUSY) partner for each bosonic
(fermionic) SM particle. It provides a qualitative understanding of various
phenomena in particle physics, as it stabilizes the gap between the Planck
and the electroweak scale \cite{Witten:1981nf}, leads to gauge coupling
unification in a straightforward way \cite{Dimopoulos:1981yj}, and includes
the lightest supersymmetric particle as a dark matter candidate
\cite{Ellis:1983ew}. Therefore the search for supersymmetric particles is
one of the main topics in the experimental program of present (Fermilab
Tevatron) and future (CERN LHC) hadron colliders.

SUSY must be broken at low energy, since spin partners of the SM particles
have not yet been observed. As a consequence, the squarks, sleptons,
charginos, neutralinos and gluino of the MSSM must be massive in comparison
to their SM counterparts. The LHC will perform a conclusive search covering
a wide range of masses up to the TeV scale. Total production cross sections
for SUSY particles at hadron colliders have been extensively studied in the
past at leading order (LO) \cite{Dawson:1983fw,delAguila:1990yw,Baer:1993ew}
and also at next-to-leading order (NLO) of perturbative QCD
\cite{Beenakker:1996ch,Beenakker:1997ut,Berger:1998kh,Berger:1999mc,%
Berger:2000iu,Baer:1997nh,Beenakker:1999xh}.

We focus our attention on slepton pair (slepton-sneutrino associated)
production at the LHC through the neutral (charged) current Drell-Yan (DY)
type processes
\begin{equation}
 \begin{array}{c}
 q \bar{q} \to \gamma,Z^0 \to \tilde{l} \tilde{l}^*, \\
 q \bar{q}' \to W^{\mp} \to \tilde{l}\tilde{\nu}_{l}^*,
                               \tilde{l}^*\tilde{\nu}_{l}.
 \end{array}
\end{equation}
Due to their purely electroweak couplings, sleptons are among the lightest
SUSY particles in many SUSY-breaking scenarios \cite{Allanach:2002nj}.
Sleptons and sneutrinos often decay directly into the stable lightest SUSY
particle (lightest neutralino in mSUGRA models or gravitino in GMSB models)
plus the corresponding SM partner (lepton or neutrino). As a result, the
slepton signal at hadron colliders will consist in a highly energetic lepton
pair, which will be easily detectable, and associated missing energy.

In this Letter, we study the transverse-momentum ($q_T$) distribution of
the slepton pair. Since in hadronic collisions the longitudinal momentum
balance is unknown, a precise knowledge of the $q_T$-balance is of vital
importance for the discovery of SUSY particles. In the case of sleptons,
the Cambridge (s)transverse mass $m_{T2}$ proves to be particularly useful
for the reconstruction of their masses \cite{Lester:1999tx} and
determination of their spin \cite{Barr:2005dz}, the two key features that
distinguish them from SM leptons produced mainly in $WW$ or $t\bar{t}$
decays \cite{Lytken:22,Andreev:2004qq}. Furthermore, both detector
kinematical acceptance and efficiency depend, of course, on $q_T$.

When studying the $q_T$-distribution of a slepton pair produced with
invariant mass $M$ in a hadronic collision, it is appropriate to separate
the large-$q_T$ and small-$q_{T}$ regions. In the
large-$q_T$ region ($q_T\geq M$) the use of fixed-order perturbation theory
is fully justified, since the perturbative series is controlled by a small
expansion parameter, $\alpha_s(M^2)$. The QCD \cite{Baer:1997nh} and full
SUSY-QCD \cite{Beenakker:1999xh} corrections for slepton pair production are
known to increase the hadronic cross sections by about 25\% at the Tevatron
and 35\% at the LHC, thus extending their discovery reaches by several tens
of GeV. Recently the LO calculation for slepton pair production has
been extended to include mixing between left- and right-handed sfermions and
longitudinal beam polarization \cite{Bozzi:2004qq}.

The bulk of the events will be produced in the small-$q_{T}$ region, where
the coefficients of the perturbative expansion in $\alpha_s(M^{2})$ are
enhanced by powers of large logarithmic terms, $\ln(M^{2}/q_{T}^{2})$. As
a consequence, results based on fixed-order calculations diverge as $q_T
\to 0$, and the convergence of the perturbative series is spoiled. These
logarithms are due to multiple soft-gluon emission from the initial state
and have to be systematically resummed to all orders in $\alpha_s$ in
order to obtain reliable perturbative predictions. The method to perform
all-order soft-gluon resummation at small $q_{T}$ is well known
\cite{Dokshitzer:1978hw,Parisi:1979se,Curci:1979bg,Kodaira:1981nh,%
Collins:1981uk,Collins:1981va,Altarelli:1984pt,Collins:1984kg,%
Davies:1984hs,Catani:2000vq,Bozzi:2005wk}. The resummation of leading logs
was first performed in \cite{Dokshitzer:1978hw}. It was shown in
\cite{Parisi:1979se} that the resummation procedure is most naturally
performed using the impact-parameter ($b$) formalism, where $b$ is the
variable conjugate to $q_T$ through a Fourier transformation, to allow the
kinematics of multiple gluon emission to factorize. In the special case of
DY lepton pair or electroweak boson production, $b$-space resummation was
performed at next-to-leading level in \cite{Kodaira:1981nh}, an all-order
resummation formalism was developed in \cite{Collins:1984kg}, and the
next-to-next-to-leading order terms have been calculated in
\cite{Davies:1984hs}.

At intermediate $q_T$ the resummed result has to be consistently matched
with fixed-order perturbation theory in order to obtain predictions with
uniform theoretical accuracy over the entire range of transverse momenta.

In this work we implement the formalism proposed in \cite{Catani:2000vq,%
Bozzi:2005wk} and compute the $q_T$-distribution of a slepton pair
produced at the LHC by combining NLL resummation at small $q_T$ and LO
(${\cal O}(\alpha_s)$) perturbation theory at large $q_T$. 

\section{\label{distribution} \boldmath$Q_T$-Resummation at the NLL level}

The partonic cross section for DY slepton pair production can be written as
\begin{equation}
\label{resplusfin}
\frac{d{\hat \sigma}_{ab}}{d M^{2} d q_T^2}=
\frac{d{\hat \sigma}_{ab}^{(\rm res.)}}{d M^{2} d q_T^2}
+\frac{d{\hat \sigma}_{ab}^{(\rm fin.)}}{d M^{2} d q_T^2}\, ,
\end{equation}
where $a,b$ label the partons which take part in the hard process. The
resummed contribution can be written as
\begin{eqnarray}
 \label{resum}
 &&\frac{d{\hat \sigma}_{ab}^{(\rm res.)}}{d M^{2} d q_T^2}(q_T,M,{\hat s};
 \alpha_s(\mu_R^2),\mu_R^2,\mu_F^2) = \\
 &&\frac{M^2}{\hat s} \int_0^\infty db \; \frac{b}{2} \;J_0(b q_T) \;
 {\cal W}_{ab}(b,M,{\hat s};\alpha_s(\mu_R^2),\mu_R^2,\mu_F^2) \;, \nonumber
\end{eqnarray}
where $J_0(x)$ is the $0^{th}$-order Bessel function, $\mu_{R}$ ($\mu_{F}$)
is the renormalization (factorization) scale, and $\hat{s}$ is the partonic
center-of-mass (CM) energy. 

The perturbative function ${\cal W}$ embodies the all-order dependence on
the large logarithms $\ln(M^2b^2)$. They correspond, in the conjugate space,
to the previously mentioned terms, $\ln(M^2/q_T^2)$, that spoil the
convergence of the perturbative series at small $q_{T}$ (large $b$).
Performing a Mellin transformation with respect to the variable
$z=M^2/{\hat s}$ at fixed $M$, we can define the $N$-moments ${\cal W}_N$ of
${\cal W}$ and express them in an exponential form
\begin{eqnarray}
 \label{wtilde}
 &&{\cal W}_N(b,M;\alpha_s(\mu_R^2),\mu_R^2,\mu_F^2)= \\
 &&{\cal H}_N\left(M, \alpha_s(\mu_R^2);M^2/\mu^2_R,M^2/\mu^2_F,M^2/Q^2
 \right)\times \nonumber\\ 
 &&\exp\{{\cal G}_N(\alpha_s(\mu^2_R),L;M^2/\mu^2_R,M^2/Q^2)\}\;, \nonumber
\end{eqnarray}
where constant ({\em i.e.} finite as $q_{T}\to0$) and logarithmically
divergent terms are factorized into the functions ${\cal H}_N$ and
${\cal G}_N$, respectively.
This factorization implies some degree of
arbitrariness, and the scale $Q$ is introduced to parameterize this
uncertainty. As in the case of $\mu_{R}$ and $\mu_{F}$, one should set
$Q=M$ and estimate the uncertainty from uncalculated subleading logarithmic
corrections by varying $Q$ around this central value. 

The function ${\cal H}_N$ does not depend on the impact parameter $b$ and,
therefore, it contains all the perturbative terms that behave as constants
in the limit $b\to\infty$. In addition it contains the whole process
dependence as well as factorization scale and scheme dependence. Its
expansion in powers of $\alpha_s$ gives
\begin{eqnarray}
\label{hexpan}
 &&{\cal H}_N(M,\alpha_s; M^2/\mu^2_R,M^2/\mu^2_F,M^2/Q^2)= \\
 &&\sigma^{(0)}(\alpha_s,M)\Big[ 1+ \frac{\alpha_s}{\pi}
 \,{\cal H}_N^{(1)}(M^2/\mu^2_R,M^2/\mu^2_F,M^2/Q^2) \nonumber \\
 && + \left( \frac{\alpha_s}{\pi} \right)^2 \,{\cal H}_N^{(2)}
 (M^2/\mu^2_R,M^2/\mu^2_F,M^2/Q^2)  + \dots\Big], \nonumber
\end{eqnarray}
where $\sigma^{(0)}$ is the lowest-order partonic cross section for the
hard-scattering process. The coefficient ${\cal H}_N^{(1)}$ splits into a
process-independent flavor off-diagonal contribution and a process-dependent
\cite{deFlorian:2000pr,deFlorian:2001zd} flavor diagonal contribution. The
general expression for ${\cal H}_{N}^{(1)}$, needed to perform a NLL
analysis, can be found in \cite{Bozzi:2005wk}. The second order coefficient
${\cal H}_{N}^{(2)}$ has not yet been computed.

The exponent $\cal G_{N}$ includes all the terms that are logarithmically
divergent when $b\to\infty$ ($q_{T}\to0$) ({\em i.e.} proportional to
$L=\ln(Q^{2}b^{2}/b_{0}^{2})$, where $b_0=2e^{-\gamma_E}$ and $\gamma_E$ is
the Euler number). However these terms become large both for small and large
$b$-values, introducing unjustified large contributions also at large
$q_T$. It is useful to introduce a modified expression of the expansion
parameter,
\begin{equation}
 \label{ldef}
 {\tilde L} \equiv \ln \left(\frac{Q^2b^2}{b_0^2} + 1\right) \ .
\end{equation}
The variables $L$ and ${\tilde L}$ are equivalent at small $q_T$, but differ
at intermediate and large $q_{T}$, avoiding the unwanted resummation
contributions since ${\tilde L}\to0$ for $Qb\ll1$ and allowing us to recover
the corresponding fixed-order total cross section after integration over
$q_T$.

The exponent ${\cal G}$ can be systematically expanded as
\begin{eqnarray}
 \label{gexpan}
 &&{\cal G}_N(\alpha_s,{\tilde L};M^2/\mu^2_R,M^2/Q^2) = \\
 &&{\tilde L}\,g^{(1)}(\alpha_s {\tilde L}) + g_N^{(2)}(\alpha_s {\tilde L};
 M^2/\mu^2_R,M^2/Q^2 )\, + \nonumber \\
 &&\frac{\alpha_s}{\pi} \;g_N^{(3)}(\alpha_s{\tilde L};M^2/\mu^2_R,M^2/Q^2 )
 + ... \ , \nonumber
\end{eqnarray}
where the term ${\tilde L} g^{(1)}$ collects the leading logarithmic (LL)
contributions, the function $g^{(2)}$ resums the next-to-leading logarithmic
(NLL) contributions, $\alpha_sg^{(3)}$ controls the next-to-next-to-leading
logarithmic (NNLL) terms, and so forth. The explicit expressions for the
$g_{i}$ functions are given in \cite{Bozzi:2005wk} in terms of the universal
perturbative coefficients $A^{(1)}_{q}$, $A^{(2)}_{q}$, $A^{(3)}_{q}$,
$B^{(1)}_{q,N}$, $B^{(2)}_{q,N}$. Since we want to perform the resummation
at the next-to-leading level, we need to implement the LL function
${\tilde L} g^{(1)}$, which depends on the coefficient $A^{(1)}_{q}$, and
the NLL function ${\tilde L} g^{(2)}$, which depends on $A^{(2)}_{q}$ and
$B^{(1)}_{q,N}$.

The second term ($d{\hat \sigma}_{ab}^{(\rm fin.)}/d M^{2} d q_T^2$) in
Eq.\ (\ref{resplusfin}) is free of divergent contributions and can be
computed by fixed-order truncation of the perturbative series. In order to
be consistently matched with the resummed contribution at intermediate
$q_T$ ($q_{T}\simeq M$), this term should be evaluated starting from the
usual perturbative truncation of the partonic cross section and subtracting
from it the expansion of the resummed part at the same perturbative order.
This matching procedure between small- and large-$q_{T}$ regions prevents
double-counting (or neglecting) of perturbative contributions and guarantees
a uniform theoretical accuracy over the entire transverse-momentum spectrum.
Since the fixed-order cross section for slepton production at non-vanishing
transverse-momentum is known at LO (slepton pair + one parton)
\cite{Baer:1997nh,Beenakker:1999xh}, we can only consistently perform a
NLL+LO matching.

The above formalism refers to a purely perturbative framework. Nonetheless
it is known \cite{Collins:1981va} that the transverse-momentum distribution
is affected by non-perturbative (NP) effects which become important in the
large-$b$ region. In the case of electroweak boson production, these
contributions are usually parameterized by multiplying the function
${\cal W}$ in Eq.\ (\ref{resum}) by a NP form factor ${\cal F}^{NP}(b)$
\cite{Davies:1984sp,Ladinsky:1993zn,Qiu:2000ga,Landry:2002ix,%
Konychev:2005iy}, whose coefficients are obtained through global fits to DY
data. We include in our analysis three different parameterizations of NP
effects corresponding to three different choices of the form factor: the
Ladinsky-Yuan (LY-G) \cite{Ladinsky:1993zn}, Brock-Landry-Nadolsky-Yuan
(BLNY) \cite{Landry:2002ix}, and the recent Konychev-Nadolsky (KN)
\cite{Konychev:2005iy} form factor.

\section{\label{results} Slepton Pair Production at the LHC}

In this Section we present quantitative results for the $q_T$-spectrum of
slepton pair (slepton-sneutrino associated) production at NLL+LO accuracy at
the LHC collider. We focus our study on the lightest slepton mass eigenstate
${\tilde \tau}_1$ and thus we consider the processes
\begin{equation}
 \begin{array}{c}
 q \bar{q} \to \gamma,Z^0 \to \tilde{\tau}_1 \tilde{\tau}^*_1, \\
 q \bar{q}' \to W^{\mp} \to \tilde{\tau}_1\tilde{\nu}_{\tau}^*,
                               \tilde{\tau}^*_1\tilde{\nu}_{\tau}.
 \end{array}
\end{equation}
We use the MRST (2004) NLO set of parton distribution functions
\cite{Martin:2004ir} and $\alpha_s$ evaluated at two-loop accuracy. We fix
the resummation scale $Q$ equal to the invariant mass $M$ of the slepton
(slepton-sneutrino) pair and we allow $\mu=\mu_{F}=\mu_{R}$ to vary between
$M/2$ and $2M$ to estimate the perturbative uncertainty. We also integrate
Eq.\ (\ref{resplusfin}) with respect to $M^{2}$, taking as lower limit the
energy threshold for $\tilde{\tau}_{1} \tilde{\tau}_{1}^{*} (\tilde{\tau}_1
\tilde{\nu}_{\tau})$ production and as upper limit the hadronic energy
($\sqrt S$=14 TeV at the LHC).

\begin{figure}
\includegraphics[scale=0.40]{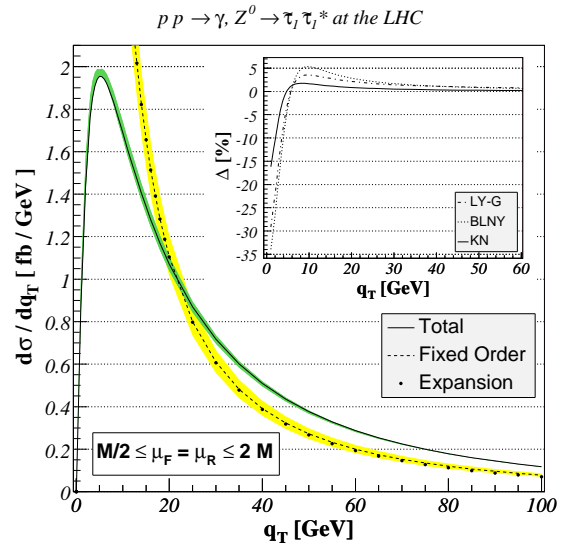}
\caption{\label{fig:Zresummation} Differential cross section for the process
 $p p \to \tilde{\tau}_{1}\tilde{\tau}_{1}^{*}$. NLL+LO matched result, LO
 result, asymptotic expansion of the resummation formula and
 $\Delta$-parameter are shown.}
\end{figure}

In the case of $\tilde{\tau}_1 \tilde{\tau}_1^*$ production (neutral
current process, see Fig.\ \ref{fig:Zresummation}), we choose the SPS7
mSUGRA benchmark point \cite{Allanach:2002nj} which gives, after the
renormalization group (RG) evolution of the SUSY-breaking parameters
performed by the SUSPECT computer program \cite{Djouadi:2002ze}, a light
${\tilde \tau}_{1}$ of mass $m_{{\tilde \tau}_{1}}=114$ GeV.   

\begin{figure}
\includegraphics[scale=0.40]{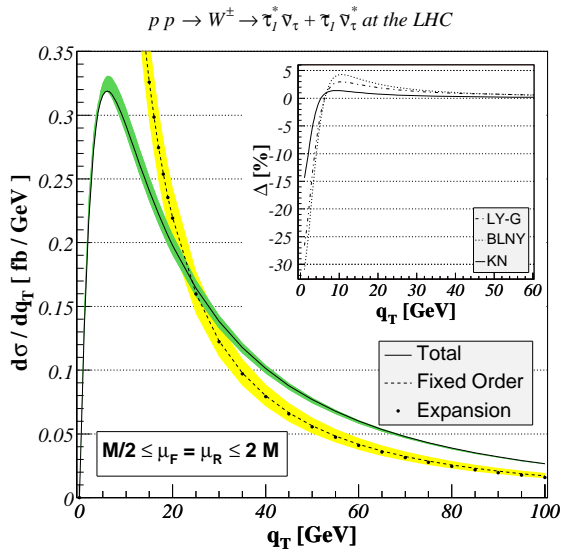}
\caption{\label{fig:Wresummation} Same as Fig.\ \ref{fig:Zresummation} for
 the process $p p \to \tilde{\tau}_1\tilde{\nu}_{\tau}^*+\tilde{\tau}_1^*
 \tilde{\nu}_{\tau}$.}
\end{figure}

In the case of $\tilde{\tau}_1\tilde{\nu}_{\tau}^*+\tilde{\tau}_1^*
\tilde{\nu}_{\tau}$ production (charged current process, see Fig.\
\ref{fig:Wresummation}), we use instead the SPS1 mSUGRA benchmark point
which gives a light ${\tilde \tau}_{1}$ of mass $m_{{\tilde \tau}_{1}}=136$
GeV as well as a light $\tilde{\nu}_{\tau}$ of mass $m_{{\tilde \nu}_{\tau}}
=196$ GeV.   

In both cases we plot the LO result (dashed line), the expansion of the
resummation formula at LO (dotted line), the total NLL+LO matched result
(solid line), the uncertainty band from scale variation, and the
quantity
\begin{equation}
 \Delta = \frac{d\sigma^{\rm (res.+NP)}(\mu=M)-d\sigma^{(\rm res.)}(\mu=M)}
 {d\sigma^{(\rm res.)} (\mu=M)}.
\end{equation}
The parameter $\Delta$ gives thus an estimate of the contributions from the
different NP parametrizations (LY-G, BLNY, KN) that we included in
the resummed formula.

We can see that the LO result diverges to $+ \infty$, as expected, for both
processes as $q_{T}\to 0$, and the asymptotic expansion of the resummation
formula at LO is in very good agreement with LO both at small and
intermediate values of $q_{T}$. The effect of resummation is clearly visible
at small and intermediate values of $q_T$, the resummation-improved result
being nearly 39\% (36\%) higher at $q_T=50$ GeV than the pure fixed order
result in the neutral (charged) current case.
When integrated over $q_T$, the former leads to a total cross section of
66.8 fb (12.9 fb) in good agreement (within 3.5\%) with the QCD-corrected
total cross section at ${\cal O}(\alpha_s)$ \cite{Beenakker:1999xh}.

The scale dependence is clearly improved in both cases with respect to the
pure fixed-order calculations. In the small and
intermediate $q_{T}$-region (up to 100 GeV) the effect of scale variation is
~10\% for the LO result, while it is always less than 5\% for the NLL+LO
curve.
Finally, non-perturbative contributions are under good control. Their effect
is always less than 5\% for $q_{T} >$ 5 GeV and thus considerably smaller
than resummation effects.

\section{\label{conclusions} Conclusions}

In this Letter, a first precision calculation of the $q_{T}$-spectrum for
SUSY particle production at hadron colliders has been performed by applying
the $q_{T}$-resummation formalism at the NLL+LO level to slepton pair and
slepton-sneutrino associated
production at the LHC. The numerical results show the importance of resummed
contributions at small and intermediate values of $q_{T}$, both enhancing
the pure fixed-order result and reducing the scale uncertainty.

\begin{acknowledgments}

We thank S.\ Catani and M.\ Grazzini for a careful reading of the
manuscript. This work was partly supported by a CNRS/IN2P3 postdoctoral
grant and a Ph.D.\ fellowship of the French ministry for education and
research.

\end{acknowledgments}

\bibliography{resum}

\begin{thebibliography}{41}
\expandafter\ifx\csname natexlab\endcsname\relax\def\natexlab#1{#1}\fi
\expandafter\ifx\csname bibnamefont\endcsname\relax
  \def\bibnamefont#1{#1}\fi
\expandafter\ifx\csname bibfnamefont\endcsname\relax
  \def\bibfnamefont#1{#1}\fi
\expandafter\ifx\csname citenamefont\endcsname\relax
  \def\citenamefont#1{#1}\fi
\expandafter\ifx\csname url\endcsname\relax
  \def\url#1{\texttt{#1}}\fi
\expandafter\ifx\csname urlprefix\endcsname\relax\def\urlprefix{URL }\fi
\providecommand{\bibinfo}[2]{#2}
\providecommand{\eprint}[2][]{\url{#2}}

\bibitem[{\citenamefont{Nilles}(1984)}]{Nilles:1983ge}
\bibinfo{author}{\bibfnamefont{H.~P.} \bibnamefont{Nilles}},
  \bibinfo{journal}{Phys. Rept.} \textbf{\bibinfo{volume}{110}},
  \bibinfo{pages}{1} (\bibinfo{year}{1984}).

\bibitem[{\citenamefont{Haber and Kane}(1985)}]{Haber:1984rc}
\bibinfo{author}{\bibfnamefont{H.~E.} \bibnamefont{Haber}} \bibnamefont{and}
  \bibinfo{author}{\bibfnamefont{G.~L.} \bibnamefont{Kane}},
  \bibinfo{journal}{Phys. Rept.} \textbf{\bibinfo{volume}{117}},
  \bibinfo{pages}{75} (\bibinfo{year}{1985}).

\bibitem[{\citenamefont{Witten}(1981)}]{Witten:1981nf}
\bibinfo{author}{\bibfnamefont{E.}~\bibnamefont{Witten}},
  \bibinfo{journal}{Nucl. Phys.} \textbf{\bibinfo{volume}{B188}},
  \bibinfo{pages}{513} (\bibinfo{year}{1981}).

\bibitem[{\citenamefont{Dimopoulos et~al.}(1981)\citenamefont{Dimopoulos, Raby,
  and Wilczek}}]{Dimopoulos:1981yj}
\bibinfo{author}{\bibfnamefont{S.}~\bibnamefont{Dimopoulos}},
  \bibinfo{author}{\bibfnamefont{S.}~\bibnamefont{Raby}}, \bibnamefont{and}
  \bibinfo{author}{\bibfnamefont{F.}~\bibnamefont{Wilczek}},
  \bibinfo{journal}{Phys. Rev.} \textbf{\bibinfo{volume}{D24}},
  \bibinfo{pages}{1681} (\bibinfo{year}{1981}).

\bibitem[{\citenamefont{Ellis et~al.}(1984)\citenamefont{Ellis, Hagelin,
  Nanopoulos, Olive, and Srednicki}}]{Ellis:1983ew}
\bibinfo{author}{\bibfnamefont{J.~R.} \bibnamefont{Ellis}},
  \bibinfo{author}{\bibfnamefont{J.~S.} \bibnamefont{Hagelin}},
  \bibinfo{author}{\bibfnamefont{D.~V.} \bibnamefont{Nanopoulos}},
  \bibinfo{author}{\bibfnamefont{K.~A.} \bibnamefont{Olive}}, \bibnamefont{and}
  \bibinfo{author}{\bibfnamefont{M.}~\bibnamefont{Srednicki}},
  \bibinfo{journal}{Nucl. Phys.} \textbf{\bibinfo{volume}{B238}},
  \bibinfo{pages}{453} (\bibinfo{year}{1984}).

\bibitem[{\citenamefont{Dawson et~al.}(1985)\citenamefont{Dawson, Eichten, and
  Quigg}}]{Dawson:1983fw}
\bibinfo{author}{\bibfnamefont{S.}~\bibnamefont{Dawson}},
  \bibinfo{author}{\bibfnamefont{E.}~\bibnamefont{Eichten}}, \bibnamefont{and}
  \bibinfo{author}{\bibfnamefont{C.}~\bibnamefont{Quigg}},
  \bibinfo{journal}{Phys. Rev.} \textbf{\bibinfo{volume}{D31}},
  \bibinfo{pages}{1581} (\bibinfo{year}{1985}).

\bibitem[{\citenamefont{del Aguila and Ametller}(1991)}]{delAguila:1990yw}
\bibinfo{author}{\bibfnamefont{F.}~\bibnamefont{del Aguila}} \bibnamefont{and}
  \bibinfo{author}{\bibfnamefont{L.}~\bibnamefont{Ametller}},
  \bibinfo{journal}{Phys. Lett.} \textbf{\bibinfo{volume}{B261}},
  \bibinfo{pages}{326} (\bibinfo{year}{1991}).

\bibitem[{\citenamefont{Baer et~al.}(1994)\citenamefont{Baer, Chen, Paige, and
  Tata}}]{Baer:1993ew}
\bibinfo{author}{\bibfnamefont{H.}~\bibnamefont{Baer}},
  \bibinfo{author}{\bibfnamefont{C.-H.} \bibnamefont{Chen}},
  \bibinfo{author}{\bibfnamefont{F.}~\bibnamefont{Paige}}, \bibnamefont{and}
  \bibinfo{author}{\bibfnamefont{X.}~\bibnamefont{Tata}},
  \bibinfo{journal}{Phys. Rev.} \textbf{\bibinfo{volume}{D49}},
  \bibinfo{pages}{3283} (\bibinfo{year}{1994}).

\bibitem[{\citenamefont{Beenakker et~al.}(1997)\citenamefont{Beenakker, Hopker,
  Spira, and Zerwas}}]{Beenakker:1996ch}
\bibinfo{author}{\bibfnamefont{W.}~\bibnamefont{Beenakker}},
  \bibinfo{author}{\bibfnamefont{R.}~\bibnamefont{Hopker}},
  \bibinfo{author}{\bibfnamefont{M.}~\bibnamefont{Spira}}, \bibnamefont{and}
  \bibinfo{author}{\bibfnamefont{P.~M.} \bibnamefont{Zerwas}},
  \bibinfo{journal}{Nucl. Phys.} \textbf{\bibinfo{volume}{B492}},
  \bibinfo{pages}{51} (\bibinfo{year}{1997}).

\bibitem[{\citenamefont{Beenakker et~al.}(1998)\citenamefont{Beenakker, Kramer,
  Plehn, Spira, and Zerwas}}]{Beenakker:1997ut}
\bibinfo{author}{\bibfnamefont{W.}~\bibnamefont{Beenakker}},
  \bibinfo{author}{\bibfnamefont{M.}~\bibnamefont{Kramer}},
  \bibinfo{author}{\bibfnamefont{T.}~\bibnamefont{Plehn}},
  \bibinfo{author}{\bibfnamefont{M.}~\bibnamefont{Spira}}, \bibnamefont{and}
  \bibinfo{author}{\bibfnamefont{P.~M.} \bibnamefont{Zerwas}},
  \bibinfo{journal}{Nucl. Phys.} \textbf{\bibinfo{volume}{B515}},
  \bibinfo{pages}{3} (\bibinfo{year}{1998}).

\bibitem[{\citenamefont{Berger et~al.}(1999{\natexlab{a}})\citenamefont{Berger,
  Klasen, and Tait}}]{Berger:1998kh}
\bibinfo{author}{\bibfnamefont{E.~L.} \bibnamefont{Berger}},
  \bibinfo{author}{\bibfnamefont{M.}~\bibnamefont{Klasen}}, \bibnamefont{and}
  \bibinfo{author}{\bibfnamefont{T.}~\bibnamefont{Tait}},
  \bibinfo{journal}{Phys. Rev.} \textbf{\bibinfo{volume}{D59}},
  \bibinfo{pages}{074024} (\bibinfo{year}{1999}{\natexlab{a}}).

\bibitem[{\citenamefont{Berger et~al.}(1999{\natexlab{b}})\citenamefont{Berger,
  Klasen, and Tait}}]{Berger:1999mc}
\bibinfo{author}{\bibfnamefont{E.~L.} \bibnamefont{Berger}},
  \bibinfo{author}{\bibfnamefont{M.}~\bibnamefont{Klasen}}, \bibnamefont{and}
  \bibinfo{author}{\bibfnamefont{T.}~\bibnamefont{Tait}},
  \bibinfo{journal}{Phys. Lett.} \textbf{\bibinfo{volume}{B459}},
  \bibinfo{pages}{165} (\bibinfo{year}{1999}{\natexlab{b}}).

\bibitem[{\citenamefont{Berger et~al.}(2000)\citenamefont{Berger, Klasen, and
  Tait}}]{Berger:2000iu}
\bibinfo{author}{\bibfnamefont{E.~L.} \bibnamefont{Berger}},
  \bibinfo{author}{\bibfnamefont{M.}~\bibnamefont{Klasen}}, \bibnamefont{and}
  \bibinfo{author}{\bibfnamefont{T.}~\bibnamefont{Tait}},
  \bibinfo{journal}{Phys. Rev.} \textbf{\bibinfo{volume}{D62}},
  \bibinfo{pages}{095014} (\bibinfo{year}{2000}).

\bibitem[{\citenamefont{Baer et~al.}(1998)\citenamefont{Baer, Harris, and
  Reno}}]{Baer:1997nh}
\bibinfo{author}{\bibfnamefont{H.}~\bibnamefont{Baer}},
  \bibinfo{author}{\bibfnamefont{B.~W.} \bibnamefont{Harris}},
  \bibnamefont{and} \bibinfo{author}{\bibfnamefont{M.~H.} \bibnamefont{Reno}},
  \bibinfo{journal}{Phys. Rev.} \textbf{\bibinfo{volume}{D57}},
  \bibinfo{pages}{5871} (\bibinfo{year}{1998}).

\bibitem[{\citenamefont{Beenakker et~al.}(1999)}]{Beenakker:1999xh}
\bibinfo{author}{\bibfnamefont{W.}~\bibnamefont{Beenakker}}
  \bibnamefont{et~al.}, \bibinfo{journal}{Phys. Rev. Lett.}
  \textbf{\bibinfo{volume}{83}}, \bibinfo{pages}{3780} (\bibinfo{year}{1999}).

\bibitem[{\citenamefont{Allanach et~al.}(2002)}]{Allanach:2002nj}
\bibinfo{author}{\bibfnamefont{B.~C.} \bibnamefont{Allanach}}
  \bibnamefont{et~al.}, \bibinfo{journal}{Eur. Phys. J.}
  \textbf{\bibinfo{volume}{C25}}, \bibinfo{pages}{113} (\bibinfo{year}{2002}).

\bibitem[{\citenamefont{Lester and Summers}(1999)}]{Lester:1999tx}
\bibinfo{author}{\bibfnamefont{C.~G.} \bibnamefont{Lester}} \bibnamefont{and}
  \bibinfo{author}{\bibfnamefont{D.~J.} \bibnamefont{Summers}},
  \bibinfo{journal}{Phys. Lett.} \textbf{\bibinfo{volume}{B463}},
  \bibinfo{pages}{99} (\bibinfo{year}{1999}).

\bibitem[{\citenamefont{Barr}(2005)}]{Barr:2005dz}
\bibinfo{author}{\bibfnamefont{A.~J.} \bibnamefont{Barr}}
  (\bibinfo{year}{2005}), \eprint{hep-ph/0511115}.

\bibitem[{\citenamefont{Lytken}(2004)}]{Lytken:22}
\bibinfo{author}{\bibfnamefont{E.}~\bibnamefont{Lytken}},
  \bibinfo{journal}{Czech. J. Phys.} \textbf{\bibinfo{volume}{54}},
  \bibinfo{pages}{A169} (\bibinfo{year}{2004}).

\bibitem[{\citenamefont{Andreev et~al.}(2005)\citenamefont{Andreev, Bityukov,
  and Krasnikov}}]{Andreev:2004qq}
\bibinfo{author}{\bibfnamefont{Y.~M.} \bibnamefont{Andreev}},
  \bibinfo{author}{\bibfnamefont{S.~I.} \bibnamefont{Bityukov}},
  \bibnamefont{and} \bibinfo{author}{\bibfnamefont{N.~V.}
  \bibnamefont{Krasnikov}}, \bibinfo{journal}{Phys. Atom. Nucl.}
  \textbf{\bibinfo{volume}{68}}, \bibinfo{pages}{340} (\bibinfo{year}{2005}).

\bibitem[{\citenamefont{Bozzi et~al.}(2005)\citenamefont{Bozzi, Fuks, and
  Klasen}}]{Bozzi:2004qq}
\bibinfo{author}{\bibfnamefont{G.}~\bibnamefont{Bozzi}},
  \bibinfo{author}{\bibfnamefont{B.}~\bibnamefont{Fuks}}, \bibnamefont{and}
  \bibinfo{author}{\bibfnamefont{M.}~\bibnamefont{Klasen}},
  \bibinfo{journal}{Phys. Lett.} \textbf{\bibinfo{volume}{B609}},
  \bibinfo{pages}{339} (\bibinfo{year}{2005}).

\bibitem[{\citenamefont{Dokshitzer et~al.}(1980)\citenamefont{Dokshitzer,
  Diakonov, and Troian}}]{Dokshitzer:1978hw}
\bibinfo{author}{\bibfnamefont{Y.~L.} \bibnamefont{Dokshitzer}},
  \bibinfo{author}{\bibfnamefont{D.}~\bibnamefont{Diakonov}}, \bibnamefont{and}
  \bibinfo{author}{\bibfnamefont{S.~I.} \bibnamefont{Troian}},
  \bibinfo{journal}{Phys. Rept.} \textbf{\bibinfo{volume}{58}},
  \bibinfo{pages}{269} (\bibinfo{year}{1980}).

\bibitem[{\citenamefont{Parisi and Petronzio}(1979)}]{Parisi:1979se}
\bibinfo{author}{\bibfnamefont{G.}~\bibnamefont{Parisi}} \bibnamefont{and}
  \bibinfo{author}{\bibfnamefont{R.}~\bibnamefont{Petronzio}},
  \bibinfo{journal}{Nucl. Phys.} \textbf{\bibinfo{volume}{B154}},
  \bibinfo{pages}{427} (\bibinfo{year}{1979}).

\bibitem[{\citenamefont{Curci et~al.}(1979)\citenamefont{Curci, Greco, and
  Srivastava}}]{Curci:1979bg}
\bibinfo{author}{\bibfnamefont{G.}~\bibnamefont{Curci}},
  \bibinfo{author}{\bibfnamefont{M.}~\bibnamefont{Greco}}, \bibnamefont{and}
  \bibinfo{author}{\bibfnamefont{Y.}~\bibnamefont{Srivastava}},
  \bibinfo{journal}{Nucl. Phys.} \textbf{\bibinfo{volume}{B159}},
  \bibinfo{pages}{451} (\bibinfo{year}{1979}).

\bibitem[{\citenamefont{Kodaira and Trentadue}(1982)}]{Kodaira:1981nh}
\bibinfo{author}{\bibfnamefont{J.}~\bibnamefont{Kodaira}} \bibnamefont{and}
  \bibinfo{author}{\bibfnamefont{L.}~\bibnamefont{Trentadue}},
  \bibinfo{journal}{Phys. Lett.} \textbf{\bibinfo{volume}{B112}},
  \bibinfo{pages}{66} (\bibinfo{year}{1982}).

\bibitem[{\citenamefont{Collins and Soper}(1981)}]{Collins:1981uk}
\bibinfo{author}{\bibfnamefont{J.~C.} \bibnamefont{Collins}} \bibnamefont{and}
  \bibinfo{author}{\bibfnamefont{D.~E.} \bibnamefont{Soper}},
  \bibinfo{journal}{Nucl. Phys.} \textbf{\bibinfo{volume}{B193}},
  \bibinfo{pages}{381} (\bibinfo{year}{1981}).

\bibitem[{\citenamefont{Collins and Soper}(1982)}]{Collins:1981va}
\bibinfo{author}{\bibfnamefont{J.~C.} \bibnamefont{Collins}} \bibnamefont{and}
  \bibinfo{author}{\bibfnamefont{D.~E.} \bibnamefont{Soper}},
  \bibinfo{journal}{Nucl. Phys.} \textbf{\bibinfo{volume}{B197}},
  \bibinfo{pages}{446} (\bibinfo{year}{1982}).

\bibitem[{\citenamefont{Altarelli et~al.}(1984)\citenamefont{Altarelli, Ellis,
  Greco, and Martinelli}}]{Altarelli:1984pt}
\bibinfo{author}{\bibfnamefont{G.}~\bibnamefont{Altarelli}},
  \bibinfo{author}{\bibfnamefont{R.~K.} \bibnamefont{Ellis}},
  \bibinfo{author}{\bibfnamefont{M.}~\bibnamefont{Greco}}, \bibnamefont{and}
  \bibinfo{author}{\bibfnamefont{G.}~\bibnamefont{Martinelli}},
  \bibinfo{journal}{Nucl. Phys.} \textbf{\bibinfo{volume}{B246}},
  \bibinfo{pages}{12} (\bibinfo{year}{1984}).

\bibitem[{\citenamefont{Collins et~al.}(1985)\citenamefont{Collins, Soper, and
  Sterman}}]{Collins:1984kg}
\bibinfo{author}{\bibfnamefont{J.~C.} \bibnamefont{Collins}},
  \bibinfo{author}{\bibfnamefont{D.~E.} \bibnamefont{Soper}}, \bibnamefont{and}
  \bibinfo{author}{\bibfnamefont{G.}~\bibnamefont{Sterman}},
  \bibinfo{journal}{Nucl. Phys.} \textbf{\bibinfo{volume}{B250}},
  \bibinfo{pages}{199} (\bibinfo{year}{1985}).

\bibitem[{\citenamefont{Davies and Stirling}(1984)}]{Davies:1984hs}
\bibinfo{author}{\bibfnamefont{C.~T.~H.} \bibnamefont{Davies}}
  \bibnamefont{and} \bibinfo{author}{\bibfnamefont{W.~J.}
  \bibnamefont{Stirling}}, \bibinfo{journal}{Nucl. Phys.}
  \textbf{\bibinfo{volume}{B244}}, \bibinfo{pages}{337} (\bibinfo{year}{1984}).

\bibitem[{\citenamefont{Catani et~al.}(2001)\citenamefont{Catani, de~Florian,
  and Grazzini}}]{Catani:2000vq}
\bibinfo{author}{\bibfnamefont{S.}~\bibnamefont{Catani}},
  \bibinfo{author}{\bibfnamefont{D.}~\bibnamefont{de~Florian}},
  \bibnamefont{and} \bibinfo{author}{\bibfnamefont{M.}~\bibnamefont{Grazzini}},
  \bibinfo{journal}{Nucl. Phys.} \textbf{\bibinfo{volume}{B596}},
  \bibinfo{pages}{299} (\bibinfo{year}{2001}).

\bibitem[{\citenamefont{Bozzi et~al.}(2006)\citenamefont{Bozzi, Catani,
  de~Florian, and Grazzini}}]{Bozzi:2005wk}
\bibinfo{author}{\bibfnamefont{G.}~\bibnamefont{Bozzi}},
  \bibinfo{author}{\bibfnamefont{S.}~\bibnamefont{Catani}},
  \bibinfo{author}{\bibfnamefont{D.}~\bibnamefont{de~Florian}},
  \bibnamefont{and} \bibinfo{author}{\bibfnamefont{M.}~\bibnamefont{Grazzini}},
  \bibinfo{journal}{Nucl. Phys.} \textbf{\bibinfo{volume}{B737}},
  \bibinfo{pages}{73} (\bibinfo{year}{2006}).

\bibitem[{\citenamefont{de~Florian and Grazzini}(2000)}]{deFlorian:2000pr}
\bibinfo{author}{\bibfnamefont{D.}~\bibnamefont{de~Florian}} \bibnamefont{and}
  \bibinfo{author}{\bibfnamefont{M.}~\bibnamefont{Grazzini}},
  \bibinfo{journal}{Phys. Rev. Lett.} \textbf{\bibinfo{volume}{85}},
  \bibinfo{pages}{4678} (\bibinfo{year}{2000}).

\bibitem[{\citenamefont{de~Florian and Grazzini}(2001)}]{deFlorian:2001zd}
\bibinfo{author}{\bibfnamefont{D.}~\bibnamefont{de~Florian}} \bibnamefont{and}
  \bibinfo{author}{\bibfnamefont{M.}~\bibnamefont{Grazzini}},
  \bibinfo{journal}{Nucl. Phys.} \textbf{\bibinfo{volume}{B616}},
  \bibinfo{pages}{247} (\bibinfo{year}{2001}).

\bibitem[{\citenamefont{Davies et~al.}(1985)\citenamefont{Davies, Webber, and
  Stirling}}]{Davies:1984sp}
\bibinfo{author}{\bibfnamefont{C.~T.~H.} \bibnamefont{Davies}},
  \bibinfo{author}{\bibfnamefont{B.~R.} \bibnamefont{Webber}},
  \bibnamefont{and} \bibinfo{author}{\bibfnamefont{W.~J.}
  \bibnamefont{Stirling}}, \bibinfo{journal}{Nucl. Phys.}
  \textbf{\bibinfo{volume}{B256}}, \bibinfo{pages}{413} (\bibinfo{year}{1985}).

\bibitem[{\citenamefont{Ladinsky and Yuan}(1994)}]{Ladinsky:1993zn}
\bibinfo{author}{\bibfnamefont{G.~A.} \bibnamefont{Ladinsky}} \bibnamefont{and}
  \bibinfo{author}{\bibfnamefont{C.~P.} \bibnamefont{Yuan}},
  \bibinfo{journal}{Phys. Rev.} \textbf{\bibinfo{volume}{D50}},
  \bibinfo{pages}{4239} (\bibinfo{year}{1994}).

\bibitem[{\citenamefont{Qiu and Zhang}(2001)}]{Qiu:2000ga}
\bibinfo{author}{\bibfnamefont{J.-W.} \bibnamefont{Qiu}} \bibnamefont{and}
  \bibinfo{author}{\bibfnamefont{X.-F.} \bibnamefont{Zhang}},
  \bibinfo{journal}{Phys. Rev. Lett.} \textbf{\bibinfo{volume}{86}},
  \bibinfo{pages}{2724} (\bibinfo{year}{2001}).

\bibitem[{\citenamefont{Landry et~al.}(2003)\citenamefont{Landry, Brock,
  Nadolsky, and Yuan}}]{Landry:2002ix}
\bibinfo{author}{\bibfnamefont{F.}~\bibnamefont{Landry}},
  \bibinfo{author}{\bibfnamefont{R.}~\bibnamefont{Brock}},
  \bibinfo{author}{\bibfnamefont{P.~M.} \bibnamefont{Nadolsky}},
  \bibnamefont{and} \bibinfo{author}{\bibfnamefont{C.~P.} \bibnamefont{Yuan}},
  \bibinfo{journal}{Phys. Rev.} \textbf{\bibinfo{volume}{D67}},
  \bibinfo{pages}{073016} (\bibinfo{year}{2003}).

\bibitem[{\citenamefont{Konychev and Nadolsky}(2006)}]{Konychev:2005iy}
\bibinfo{author}{\bibfnamefont{A.~V.} \bibnamefont{Konychev}} \bibnamefont{and}
  \bibinfo{author}{\bibfnamefont{P.~M.} \bibnamefont{Nadolsky}},
  \bibinfo{journal}{Phys. Lett.} \textbf{\bibinfo{volume}{B633}},
  \bibinfo{pages}{710} (\bibinfo{year}{2006}).

\bibitem[{\citenamefont{Martin et~al.}(2004)\citenamefont{Martin, Roberts,
  Stirling, and Thorne}}]{Martin:2004ir}
\bibinfo{author}{\bibfnamefont{A.~D.} \bibnamefont{Martin}},
  \bibinfo{author}{\bibfnamefont{R.~G.} \bibnamefont{Roberts}},
  \bibinfo{author}{\bibfnamefont{W.~J.} \bibnamefont{Stirling}},
  \bibnamefont{and} \bibinfo{author}{\bibfnamefont{R.~S.}
  \bibnamefont{Thorne}}, \bibinfo{journal}{Phys. Lett.}
  \textbf{\bibinfo{volume}{B604}}, \bibinfo{pages}{61} (\bibinfo{year}{2004}).

\bibitem[{\citenamefont{Djouadi et~al.}(2002)\citenamefont{Djouadi, Kneur, and
  Moultaka}}]{Djouadi:2002ze}
\bibinfo{author}{\bibfnamefont{A.}~\bibnamefont{Djouadi}},
  \bibinfo{author}{\bibfnamefont{J.-L.} \bibnamefont{Kneur}}, \bibnamefont{and}
  \bibinfo{author}{\bibfnamefont{G.}~\bibnamefont{Moultaka}}
  (\bibinfo{year}{2002}), \eprint{hep-ph/0211331}.

\end{thebibliography}

\end{document}